\documentclass[11pt]{article}
\usepackage{amsfonts}
\usepackage{amssymb}
\usepackage{color}
\usepackage{graphicx}
\makeatletter
  \@addtoreset{equation}{section}
  \renewcommand{\theequation}{\thesection.\@arabic\c@equation}
  \makeatother

\begin{document}

\title{A Combinatorial Optimization Approach to the Stability of Biomacromolecular Structures}
% Additionally, use \titlerunning for an abbreviated version of
% your title (only if the full title is too long to be displayed
% in the header line of the following pages), else comment out

\author{{\bf {\large R. P. Mondaini}}  \\
 {\small Federal University of Rio de Janeiro - UFRJ,} \\
 {\small Technology Centre, COPPE,}\\
 {\small 21941-972, Rio de Janeiro, RJ, Brazil, P.O. Box 68511.} \\
 {\small {\tt mondaini@cos.ufrj.br, rpmondaini@gmail.com}.}{\small}}

\date{}
\maketitle

\maketitle

\begin{abstract}

The application of optimization techniques derived from the study of
Euclidean full Steiner Trees to macromolecules like proteins is
reported in the present work. We shall use the concept of Euclidean
Steiner Ratio Function (SRF) as a good Lyapunov function in order to
perform an elementary stability analysis.

{\bf Keywords:} Steiner, biomacromolecular structure, full trees,
geometric chirality.
% please provide a list of keywords for your work
\end{abstract}

%%%%%%%%%%%%%%%%%%%%%%%%%%%%%%%%%%%%%%%%%%%%%%%%%%%%%%%%%%%%%%%%%%%

\section{Introduction}
\label{sec:1} Nature has followed mathematical principles of
structural organization in the construction of macromolecular
configurations. Our proposal in the present work is the modelling of
the folded stage of proteins by some combinatorial optimization
techniques associated to Euclidean full Steiner trees \cite{SMIM}.
This means that henceforth we take the 3-dimensional Euclidean space
$\mathbb{E}^3$ as our metric manifold. The analysis to be undertaken
can be summarized by the trial of obtaining the potential energy
minimization of a protein structure through the problem of length
minimization of an Euclidean Steiner Tree \cite{rm:gmb,MON3}. Our
fundamental pattern of input points will be given by sets of evenly
spaced points along a right circular helix of unit radius. We have,
\begin{equation}
P_j=(\cos j\omega, \sin j\omega, \alpha j\omega);
\,\,\,\,\,\,\,0\leq j \leq n-1,
\end{equation}
where $\omega$ is the angular coordinate and $2\pi\alpha$ stands for
the pitch of the helix.

We also use the result of Steiner points belonging to another helix
of the same pitch and smaller radius or
\begin{equation}
S_k=(r(\omega,\alpha)\cos k\omega, r(\omega,\alpha)\sin k\omega,
\alpha k\omega);\,\,\,\,\,\,\,1\leq k \leq n-2,
\end{equation}
where
\begin{equation}
r(\omega,\alpha)=\frac{\alpha\omega}{\sqrt{A_1(A_1+1)}};\,\,\,\,\,\,\,A_1=1-2\cos
\omega.
\end{equation}

The function $r(\omega,\alpha)$ above is easily obtained from the
requirement of meeting edges at an angle of $120^{\mathrm o}$ on
each Steiner point. To be rigorous, we should write,

\begin{equation}
r(\omega,\alpha)=\mbox{Max}\left(1,\frac{\alpha\omega}{\sqrt{A_1(A_1+1)}}\right),
\end{equation}

where the Max above should be understood as a piecewise choice of
the Maple$\circledR$ software.

\section{Trees of Helical Point Sets}

We now introduce a generalization of the formulae above by thinking
on subsequences of input and Steiner points, corresponding to
non-consecutive points. These subsequences are of the form:
\begin{equation}
(P_j)_{m,\,l_{P max}}: P_j, P_{j+m}, P_{j+2m},\ldots,P_{j+l_{P
max}.m}
\end{equation}
\begin{equation}
(S_k)_{m,\,l_{S max}}: S_k, S_{k+m}, S_{k+2m},\ldots,S_{k+l_{S
max}.m}
\end{equation}
where $l_P$, $l_S$ are the number of intervals of skipped points
before the present point on each subsequence and $(m-1)$ is the
number of skipped points.

We also have:
\begin{eqnarray}
0\leq l_P\leq l_{P max} =
\left[\frac{n-j-1}{m}\right];\,\,\,\,\,\,\,\,\,1\leq l_S\leq l_{S
max} =\left[\frac{n-k-2}{m}\right]\\ \nonumber\\0\leq j\leq m-1,
\,\,\,\,\,\,\,1\leq k \leq m \nonumber
\end{eqnarray}
and the square brackets stand for the greatest integer value.

The sequences corresponding to eqs.(1.1) and (1.2) are of course
included in the scheme above. They are $(P_0)_{1,n-1}$ and
$(S_1)_{1,n-2}$, respectively. In the general case, we can define
new sequences of $n$ and $n-2$ points instead those given by
eqs.(1.1) and (1.2). We shall have respectively,
\begin{equation}
{\mathbb P}_m=\bigcup_{j=0}^{m-1}(P_j)_{m,l_{j
max}};\,\,\,\,\,\,\,\,\,{\mathbb S}_m=\bigcup_{k=1}^{m}(S_k)_{m,l_{k
max}}.
\end{equation}

The present development is independent of a specific coordinate
representation of the points. If we now assume helical point sets
whose points are evenly spaced along right circular helices, we get
\begin{equation}
P_{j+l_P m}=(\cos (j+l_P m)\omega, \sin(j+l_P m)\omega, \alpha(j+l_P
m)\omega),
\end{equation}
\begin{equation}
S_{k+l_S m}=(r_m(\omega,\alpha)\cos (k+l_S m)\omega,
r_m(\omega,\alpha)\sin (k+l_S m)\omega, \alpha(k+l_S m)\omega).
\end{equation}

The function $r_m(\omega,\alpha)$ is obtained through the same
requirement of meeting edges at $120^{\mathrm o}$ on each Steiner
point. We have, analogously,
\begin{equation}
r_m(\omega,\alpha)=\mbox{Max}\left(1,\frac{m\alpha\omega}{\sqrt{A_m(A_m+1)}}\right),
\end{equation}
where
\begin{equation}
A_m=1-2\cos (m\omega).
\end{equation}

In figure (1) below we show some sequences of input points for
$n=23$.\pagebreak

\begin{figure} [hp]
\begin{center}
\includegraphics[width=1.15\textwidth,bb=0 0 640 480]{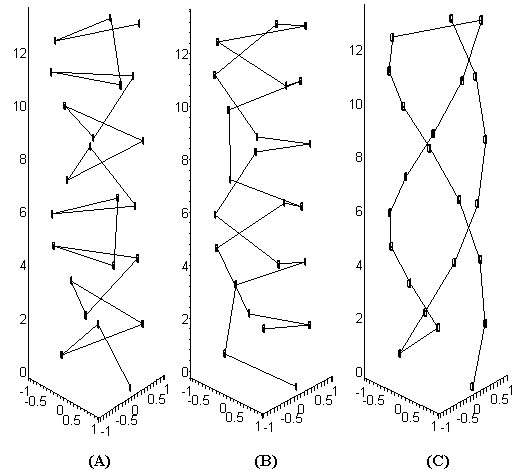}
\end{center}
\caption{\small{{\bf (A)} The sequence $n=23$, $m=1$, $j=0$; {\bf
(B)} The union of the subsequences $n=23$, $m=2$, $j=0$ and $n=23$,
$m=2$, $j=1$; {\bf (C)} The union of the subsequences $n=23$, $m=3$,
$j=0$; $n=23$, $m=3$, $j=1$ and $n=23$, $m=3$, $j=2$.}} \label{f1}
\end{figure}

From eqs.(2.5) and (2.6) and figure (1), we can write for the length
of the spanning trees
\begin{equation}
l_{SP}(m, \omega,\alpha)=\sqrt{m^2\alpha^2\omega^2
+A_m+1}\,\,\,.\,\,\sum_{j=0}^{m-1}\left[\frac{n-j-1}{m}\right]+(m-1)\sqrt{\alpha^2\omega^2+A_1+1}.
\end{equation}

The length of the Steiner Trees is then
\begin{eqnarray}
l_{ST}(m,\omega,\alpha)&=&(1-r_m(\omega,\alpha))\left(m+\sum_{k=1}^{m}\left[\frac{n-k-2}{m}\right]\right)\\
&+&\sqrt{m^2\alpha^2\omega^2+r_m(\omega,\alpha)(A_m+1)}\,\,\,.\,\,\sum_{k=1}^{m}\left[\frac{n-k-2}{m}\right]\nonumber\\
&+&2\sqrt{m^2\alpha^2\omega^2+(1-r_m(\omega,\alpha))^2+r_m(\omega,\alpha)(A_m+1)}\nonumber.
\end{eqnarray}

After using some useful relations like
\begin{equation}
\sum_{j=0}^{m-1}\left[\frac{n-j-1}{m}\right]=n-m
\end{equation}
\begin{equation}
\sum_{k=1}^{m}\left[\frac{n-k-2}{m}\right]=n-m-2
\end{equation}
and taking the limit for $n\gg m$, we get
\begin{equation}
l_{SP}(m,\omega,\alpha)=n\sqrt{m^2\alpha^2\omega^2+A_m+1}
\end{equation}
\begin{equation}
l_{ST}(m,\omega,\alpha)=n\left(1+m\alpha\omega\sqrt{\frac{A_m}{A_m+1}}\right).
\end{equation}

By following the prescriptions for writing the Steiner Ratio, we can
write for the Steiner Ratio Function of very large helical point
sets with points evenly spaced along right circular helices
\begin{equation}
\rho(\omega,\alpha)=\frac{\min_{(m)}\left(1+m\alpha\omega\sqrt{\frac{A_m}{A_m+1}}\right)}{\min_{(m)}\left(\sqrt{m^2\alpha^2\omega^2+A_m+1}\right)}
\end{equation}
where the $\min$ process above should be understood in the sense of
a piecewise function formed by the functions corresponding to the
values $m=1,2,3,\ldots$

Eq.(2.15) is our proposal for a Steiner Ratio Function (SRF)
\cite{rm:arxiv1,rm:arxiv2}. It allows for an analytic formulation of
the search of the Steiner Ratio which is then defined as the minimum
of the SRF function, eq.(2.15). Actually, there is a further
restriction to be imposed on function (2.15) in order to
characterize it as an useful SRF function. This restriction is that
we should consider only full Steiner Trees, i.e., non-degenerated
Steiner trees in which there are exactly three edges meeting at each
Steiner point. This restriction can be imposed on the spanning
trees, by requesting that the angle $\theta_m(\omega,\alpha)$
between consecutive edges formed with the points $P_{j+l_P m}$ as
vertices should be lesser than $120^{\mathrm o}$. We have
\begin{equation}
-\frac{1}{2}\leq\cos
\theta_m(\omega,\alpha)=-1+\frac{(A_m+1)^2}{2(m^2\alpha^2\omega^2+A_m+1)}.
\end{equation}

In figure (2) below we can see the restrictions corresponding to
eq.(2.16), for $m=1,2,3$. The horizontal line is $\cos
\theta_m=-1/2$.

\begin{figure} [hp]
\hfil\scalebox{0.9}{\includegraphics{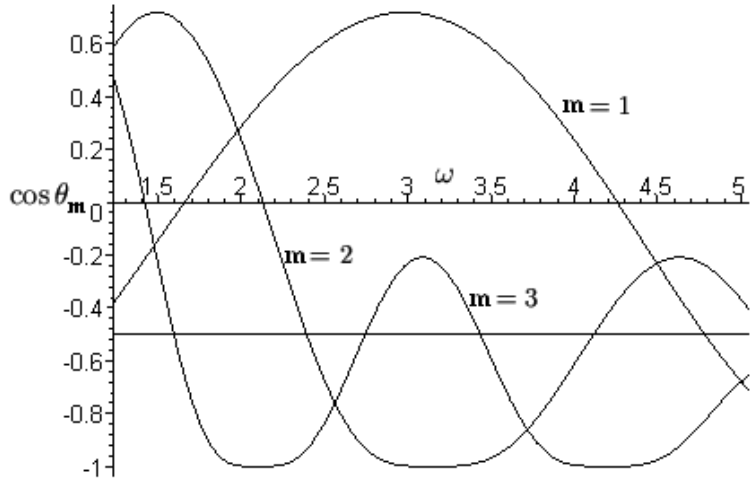}}\hfil \caption{\small
The restriction to Full Steiner Trees. The figure is a section
$\alpha=\alpha_R$ (eq.(2.19)) of the surfaces given by eq.(2.16)
corresponding to $m=1,2,3$. } \label{f2}
\end{figure}

The $m=1$ spanning tree is the only one which corresponds to Full
Steiner trees in a large region of the $\omega$-interval convenient
for our work. The other trees, $m=2,3$ correspond to forbidden
regions in the same $\omega$-interval. The corresponding Steiner
trees to be obtained from the positions of the points $S_{k+l_S m}$
and $P_{j+l_P m}$ are necessarily degenerate and should not be taken
into consideration. Thus, the prescription (2.15) for the SRF
function turns into
\begin{equation}
\rho(\omega,\alpha)=\frac{1+\alpha\omega\sqrt{\frac{A_1}{A_1+1}}}{\min_{(m)}\left(\sqrt{m^2\alpha^2\omega^2+A_m+1}\right)}=\mbox{Max}_{(m)}\,\,\rho_m(\omega,\alpha)
\end{equation}
where
\begin{equation}
\rho_m(\omega,\alpha)=\frac{1+\alpha\omega\sqrt{\frac{A_1}{A_1+1}}}{\sqrt{m^2\alpha^2\omega^2+A_m+1}}.
\end{equation}

The function (2.17) has a global minimum in the point
\begin{equation}
(\omega_R,\alpha_R)=\left(\pi-\arccos \frac{2}{3},
\frac{\sqrt{30}}{9\left(\pi-\arccos\frac{2}{3}\right)}\right)
\end{equation}
and
\begin{equation}
\rho(\omega_R,\alpha_R)=\frac{1}{10}(3\sqrt{3}+\sqrt{7})=0.78419037337\ldots
\end{equation}

For a proof see \cite{rm:arxiv1}.

The last value corresponds to the famous main conjecture of ref.
\cite{SMIM} about the value of the Steiner Ratio in 3-dimensional
Euclidean Space. It lead us also to think that Nature has solved the
problem of energy minimization in the organization of intramolecular
structure by choosing Steiner Trees as an intrinsic part of this
structure \cite{ms:eucl}.

\section{The Stability of Steiner Trees Under Elastic Force Deformation}
\label{sec:stab}

In the following we continue to work in a ${\mathbb R}^3$ manifold
with an Euclidean distance definition. Let us now introduce a tree
as that of figure (3)

\begin{figure} [hp]
\hfil\scalebox{0.5}{\includegraphics{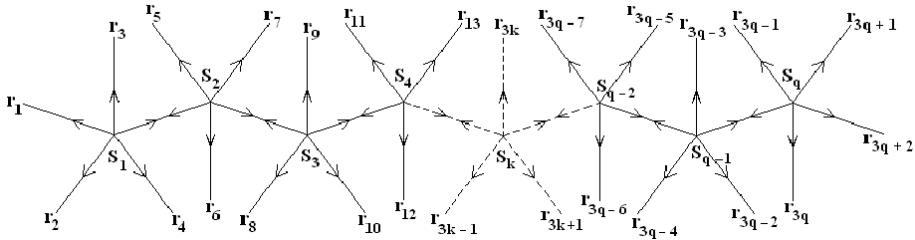}}\hfil \caption{\small
Geometrical scheme for a Steiner Problem with $p=5$.} \label{f3}
\end{figure}

There are $n$ input points (position vectors $\mathbf{r}_j$) and
$q=(n-2)/(p-2)$ Steiner points (position vectors $\mathbf{S}_k$). If
$q$ is not an integer number, there is not a tree with these $n$,
$p$ values \cite{MON1}. In figure (3) with $p=5$, we assume $n$ to
be a feasible value. The knowledge of the Steiner Problem tell us
that this tree structure is not stable since its total length can be
reduced by decreasing the number $p$ \cite{GIL}. The usual stable
Steiner problem corresponds to $p=3$. In this section we shall give
another proof of this fact by exploring the concept of a Steiner
network with physical interaction among its vertices. The structure
depicted at figure (3) is a representative of the network which
models the fundamental interactions inside a biomacromolecule. Let
us consider the interaction of this structure with similar
structures. Let the resulting interaction forces as applied to input
and Steiner points be $\mathbf{f}_j$, $\mathbf{f}_{S_k}$,
respectively and let $l_{S_k r_j}$ be the length of an edge between
a Steiner point and an input point on its neighbourhood. We have the
following identities:
\begin{equation}
l_{S_k
r_j}=\frac{1}{a_j}\,\mathbf{f}_j\,\,\cdot\,\,(\mathbf{r}_j-\mathbf{f}_{S_k})
\end{equation}
\begin{equation}
l_{S_k
S_{k+1}}=\frac{1}{a_{S_k}}\,\mathbf{f}_{S_k}\,\,.\,\,(\mathbf{S}_{k+1}-\mathbf{S}_k)
\end{equation}
where $a_j$, $a_{S_k}$, $j=1,2,\ldots,n$, $k=1,2,\ldots,q$ stand for
the modulus of the parallel components to the edges of the resulting
forces $\mathbf{f}_j$, $\mathbf{f}_{S_k}$, respectively.

The total length of the tree above is
\begin{equation}
l=\sum_{j=1}^{m-1}l_{S_1 r_j}+\sum_{j=m}^{2m-3}l_{S_2
r_j}+\sum_{j=2m-2}^{3m-5}\!\!\!l_{S_3
r_j}+\ldots+\sum_{j=n-m+2}^{n}\!\!\!\!\!l_{S_q
r_j}+\,\,\sum_{k=1}^{q-1}l_{S_k S_{k+1}}
\end{equation}

From eqs.(2.20), (3.1), we can write the total length in the form
\begin{eqnarray}
l&=&\sum_{j=1}^{n}\mathbf{r}_j\,\,\cdot\,\,\frac{\mathbf{f}_j}{a_j}-\mathbf{S}_1\cdot\left(\frac{\mathbf{f}_{S_1}}{a_1}+\sum_{j=1}^{p-1}\frac{\mathbf{f}_j}{a_j}\right)\nonumber\\
&-&\sum_{k=2}^{q-1}\mathbf{S}_k\,\,\cdot\sum_{j=(k-1)p-2k+4}^{kp-2k+1}\left(\frac{\mathbf{f}_j}{a_j}+\frac{\mathbf{f}_{S_k}}{a_{S_k}}+\frac{(-\mathbf{f}_{S_{k-1}})}{a_{S_{k-1}}}\right)\nonumber\\
&-&\mathbf{S}_q\,\,\cdot\,\left(\frac{(-\mathbf{f}_{S_{q-1}})}{a_{S_{q-1}}}+\sum_{j=n-p+2}^{n}\frac{\mathbf{f}_j}{a_j}\right)\,.
\end{eqnarray}

We now specialize this set of applied forces at the vertices as
being collinear with the edges joining them, or
\begin{equation}
\mathbf{f}_j=a_j
\hat{f}_{j||}\,\,;\,\,\,\,\,\,\,\,\,\,\,\,\,\,\mathbf{f}_{S_k}=a_{S_k}\hat{f}_{S_k
||}
\end{equation}
where the double vertical stroke means ``collinear with the edge"
and the hat over a letter stands as unit vector.

We now assume that the forces along the edges are Hooke elastic
forces
\begin{equation}
\mathbf{f}_{j
||}=C(\mathbf{r}_j-\mathbf{S}_k)\,\,\,\,\,\,\,\,\,\mbox{or}\,\,\,\,\,\,\,\,\,\hat{f}_{j||}=\hat{r}_j=\frac{\mathbf{r}_j-\mathbf{S}_k}{||\mathbf{r}_j-\mathbf{S}_k||}
\end{equation}
\begin{equation}
\mathbf{f}_{S_k
||}=C(\mathbf{S}_{k+1}-\mathbf{S_k})\,\,\,\,\,\,\,\,\,\mbox{or}\,\,\,\,\,\,\,\,\,\hat{f}_{S_k
||}=\hat{S}_k=\frac{\mathbf{S}_{k+1}-\mathbf{S}_k}{||\mathbf{S}_{k+1}-\mathbf{S}_k||}
\end{equation}
where $C$ is the elastic constant.

The assumption of local equilibrium of these forces lead to the
conditions:
\begin{equation}
\sum_{j=1}^{p-1}\hat{r}_j+\hat{S}_1=0
\end{equation}
\begin{equation}
\sum_{j=(k-1)p-2k+4}^{kp-2k+1}\hat{r}_j+\hat{S}_k-\hat{S}_{k-1}=0,
\,\,\,\,\,\,\,\,\,\,k=2,3,\ldots,q-1
\end{equation}
\begin{equation}
\sum_{j=n-p+2}^{n}\hat{r}_j-\hat{S}_{q-1}=0\,.
\end{equation}
This is a set of generalized Fermat problems or Steiner Problems
\cite{hw:kuhn}.

For this equilibrium configuration, eq.(3.3) turns into
\begin{equation}
l=\sum_{j=1}^{n}\mathbf{r}_j\,\,\cdot\,\,\hat{f}_{j ||}\,.
\end{equation}

The stability of this equilibrium configuration under a variation of
the applied forces can be tested by
\begin{equation}
\delta l=\sum_{j=1}^{n}\mathbf{r}_j\,\,\cdot\,\,\delta\hat{f}_{j
||}=0\,.
\end{equation}

We take cartesian coordinates for the ${\mathbb R}^3$ vectors
$\mathbf{r}_j=(x_i, y_i, z_i)$, $\mathbf{S}_k=(x_{S_k}, y_{S_k},
z_{S_k})$ and we consider the three independent variations
$\delta_{x_{S_k}}$, $\delta_{y_{S_k}}$, $\delta_{z_{S_k}}$ in the
coordinates of the Steiner points.

The corresponding variations in the length of the tree are of the
form
\begin{equation}
\delta_{x_{S_k}}l=\delta_{x_{S_k}}\cdot\sum_{j=1}^{n}\frac{-x_j(\mathbf{r}_j-\mathbf{S}_k)^2+[(\mathbf{r}_j-\mathbf{S}_k)\cdot\mathbf{r}_j](x_j-x_{S_k})}{||\mathbf{r}_j-\mathbf{S}_k||^3}=0
\end{equation}
and two other analogous expressions for the variations
$\delta_{y_{S_k}}$, $\delta_{z_{S_k}}$.

From the arbitrariness of these variations we can write,
\begin{equation}
\sum_{j=1}^{n}\frac{(\mathbf{r}_j-\mathbf{S}_k)\times[(\mathbf{r}_j-\mathbf{S}_k)\times\mathbf{r}_j]}{||\mathbf{r}_j-\mathbf{S}_k||^3}=0\,.
\end{equation}

We can also write
\begin{equation}
\sum_{j=1}^{n}\frac{(\mathbf{r}_j\,\,\cdot\,\,\mathbf{S}_k)^2 -r_j^2
S_k^2}{||\mathbf{r}_j-\mathbf{S}_k||^3}=0\,.
\end{equation}

We now write the position vectors $\mathbf{r}_j$, $\mathbf{S}_k$ for
the configuration depicted at figure (3). The points can be taken as
evenly spaced along right circular helices which radii are 1 and
$R_p$, respectively. We have,
\begin{equation}
\mathbf{r}_j=(\cos (j-1)\omega, \sin(j-1)\omega, \alpha(j-1)\omega)
\end{equation}
\begin{equation}
\mathbf{S}_k=(R_p\cos k\omega, R_p\sin k\omega, \alpha k\omega)\,.
\end{equation}

$R_p(\omega ,\alpha)$ is a function which can be derived from the
equilibrium conditions in eqs.(3.7)--(3.9). For $p=3$ there is only
one solution given by
\begin{equation}
R_3(\omega,\alpha)=\frac{\alpha\omega}{\sqrt{A_1(A_1+1)}}\,\,;\,\,\,\,\,\,
A_1=1-2\cos \omega\,.
\end{equation}

This solution coincides with eq.(1.3).

For $p>3$, another useful solution could be obtained from the
equations:
\begin{eqnarray}
\cos(\hat{r}_j,\hat{r}_l)=\cos(\hat{r}_j,\hat{S}_k)=\cos(\hat{S}_k,\hat{S}_m)=-\frac{1}{(p-1)}\nonumber\\
\hspace{3cm}j,l=1,2,\cdots,n;\,\,\,\,k,m=1,2,\cdots,q.\nonumber
\end{eqnarray}

Curiously, Nature has chosen this solution for $p=4$ to keep sure of
partial equilibrium of side chains between the Amide plane
conformation in proteins \cite{ms:eucl,vo:bioch,gl:crys}.

For the configuration given by eqs.(3.15)--(3.16), eq.(3.15) can be
written as
\[\sum_{j=1}^{n}T_{jkp}(\omega,\alpha)=0\]
where the geometrical object $T_{jkp}$ can be written in the
coordinates of eqs.(3.16) and (3.17) as
\[T_{jkp}=\hspace{11.8cm}\]
\[\frac{[\cos^2(j\!\!-\!\!1\!\!-\!\!k)\omega\!\!-\!\!1\!\!-\!\!\alpha^2\omega^2(j\!\!-\!\!1)^2]R_p^2+\alpha^2\omega^2[2R_p(j\!\!-\!\!1)k\cos(j\!\!-\!\!1\!\!-\!\!k)\omega\!\!-\!\!k^2]}{[1+R_p^2\!\!-\!\!2R_p\cos(j\!\!-\!\!1\!\!-\!\!k)\omega+\alpha^2\omega^2 (j\!\!-\!\!1\!\!-\!\!k)^2]^{3/2}}.\]
\begin{equation}
\end{equation}

To each $k$-value, there will be a term $j=k+1$ which dominates the
sum above. However, we cannot have $j=k+1$ for $p>3$. This can be
seen from the fact for a vertex $S_k$ ($k\neq 1,q$) there are
($p-2$) nearest external vertices $r_j$. The sequence of their
consecutive position vectors is
\begin{equation}
\mathbf{r}_{(k-1)p-2k+4}, \ldots\ldots,\mathbf{r}_{kp-2k+1}
\end{equation}
and the requirement $j=k+1$ corresponds to an integer $p$-value only
for $p=3$.

This $p=3$ case which is known to correspond to the most stable
problem \cite{GIL} has as a possible configuration the figure (4)
below

\begin{figure} [hp]
\hfil\scalebox{0.65}{\includegraphics{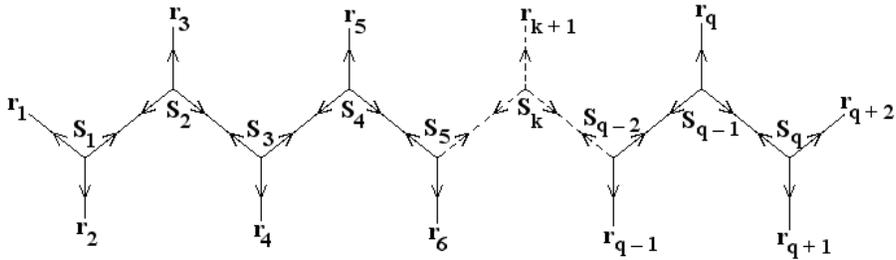}}\hfil
\caption{\small The stable structure of the $p=3$ Steiner Problem. }
\label{f4}
\end{figure}

\section{Concluding Remarks}

We have stressed on some past publications that there is a
self-consistent treatment of the intramolecular organization of
biomacromolecules in terms of Steiner networks. This representation
is able at deriving information concerning its stability and
evolution. The supporting facts for stability are now
well-established and the ideas related to the evolution of
macromolecules are in their way to be developed and accepted as a
preliminary theory of molecular evolution. The missing subject is a
full description of geometric chirality and in order to unveil some
of its properties, we have proposed to study the influence of some
proposals for chirality measure on the dynamics of optimization
problems. These are aimed at studying the structures which energy is
around the assumed energy of the minimum solution and the variation
process of the chiral properties in the neighbourhood of this
minimum. We think that this research line is worth of serious
scientific work and should take advantage of the best efforts of
very good scientific researchers for some years.

%%%%%%%%%%%%%%%%%%%%%%%%%%%%%%%%%%%%%%%%%%%%%%%%%%%%%%%%%%%%%%%%%%%%%%

\end{document}